\documentclass[10pt,conference]{IEEEtran}
\IEEEoverridecommandlockouts
\usepackage{cite}
\usepackage{amsmath,amssymb,amsfonts,mathtools}
\usepackage{algorithmic}
\usepackage{graphicx}
\usepackage{textcomp}
\usepackage{xcolor}
\def\BibTeX{{\rm B\kern-.05em{\sc i\kern-.025em b}\kern-.08em
    T\kern-.1667em\lower.7ex\hbox{E}\kern-.125emX}}

\usepackage{setspace}

\usepackage{glossaries}
\newacronym[firstplural=spatio-temporal covariance matrices (STCMs)]{STCM}{STCM}{spatio-temporal covariance matrix}
\newacronym{BLCMP}{BLCMP}{binaural linearly constrained minimum power}
\newacronym{RIR}{RIR}{room impulse response}
\newacronym{LCMP}{LCMP}{linearly constrained minimum power}
\newacronym{wBLCMP}{wBLCMP}{weighted binaural linearly constrained minimum power}
\newacronym{STFT}{STFT}{short-time Fourier transform}
\newacronym{CTF}{CTF}{convolutive transfer function}
\newacronym{MTF}{MTF}{multiplicative transfer function}
\newacronym{RTF}{RTF}{relative transfer function}
\newacronym{MCLP}{MCLP}{multi channel linear prediction}
\newacronym{MPDR}{MPDR}{minimum power distortionless response}
\newacronym{MVDR}{MVDR}{minimum variance distortionless response}
\newacronym{LCMV}{LCMV}{linear constrained minimum variance}
\newacronym{WPD}{WPD}{weighted power minimization distortionless response}
\newacronym{WPE}{WPE}{weighted prediction error}
\newacronym{TVG}{TVG}{time-varying complex circular Gaussian}
\newacronym{MISO}{MISO}{multiple-input single-output}
\newacronym{MIMO}{MIMO}{multiple-input multiple-output}
\newacronym{SPP}{SPP}{speech presence probability}
\newacronym{PESQ}{PESQ}{perceptual evaluation of speech quality}
\newacronym{FWSSNR}{FWSSNR}{frequency-weighted segmental signal-to-noise ratio}
\newacronym{CW}{CW}{covariance whitening}
\newacronym{VAD}{VAD}{voice activity detection}
\newacronym{PSD}{PSD}{power spectral density}
\newacronym{UCB}{UCB}{unified convolutional beamformer}
\newacronym{IRLS}{IRLS}{iteratively reweighted least squares}
\newacronym{MFMVDR}{MFMVDR}{multi-frame minimum variance distortionless response}
\newacronym{BMFMVDR}{BMFMVDR}{binaural MFMVDR}
\newacronym{TCN}{TCN}{temporal convolutional network}
\newacronym{DNN}{DNN}{deep neural network}
\newacronym{SNR}{SNR}{signal-to-noise ratio}
\newacronym{DRC}{DRC}{dynamic range compression}
\newacronym{HGR}{HGR}{half-gain rule}
\newacronym{CEC1}{CEC1}{Clarity Enhancement Challenge}
\newacronym{MBDRC}{MBDRC}{multi-band dynamic range compressor}
\newacronym{MBSTOI}{MBSTOI}{modified binaural short-term objective intelligibility}
\newacronym{SD-SDR}{SD-SDR}{scale-dependent signal-to-distortion ratio}
\newacronym{IFC}{IFC}{inter-frame correlation}
\newacronym{BTE}{BTE}{behind-the-ear}
\newacronym{SIR}{SIR}{signal-to-interferer ratio}
\newacronym{SRR}{SRR}{signal-to-reverberation ratio}
\newacronym{DUR}{DUR}{desired-to-undesired ratio}
\newacronym{wMPDR}{wMPDR}{weighted \gls{MPDR}}

\DeclareMathOperator*{\argmin}{argmin}
\DeclarePairedDelimiter\abs{\lvert}{\rvert}%
\DeclarePairedDelimiter\norm{\lVert}{\rVert}%
\makeatletter
\let\oldabs\abs
\def\abs{\@ifstar{\oldabs}{\oldabs*}}
\let\oldnorm\norm
\def\norm{\@ifstar{\oldnorm}{\oldnorm*}}
\makeatother

\usepackage{siunitx}
\usepackage{units}

\usepackage{tikz,pgfplots}
\usetikzlibrary{matrix}
\pgfplotsset{compat=newest, width=3.5cm, height=2.3cm, scale only axis}
\pgfplotsset{every mark/.append style={solid}}
\usepackage{pgfkeys}
    \newenvironment{customlegend}[1][]{%
        \begingroup
        \csname pgfplots@init@cleared@structures\endcsname
        \pgfplotsset{#1}%
    }{%
        \csname pgfplots@createlegend\endcsname
        \endgroup
    }%
    \def\addlegendimage{\csname pgfplots@addlegendimage\endcsname}

\usepackage{balance}
\usepackage{lastpage}

\usepackage{tikz}
\usepackage{environ}
\makeatletter
\newsavebox{\measure@tikzpicture}
\NewEnviron{scaletikzpicturetowidth}[1]{%
  \def\tikz@width{#1}%
  \begin{lrbox}{\measure@tikzpicture}%
  \BODY
  \end{lrbox}%
  \pgfmathparse{#1/\wd\measure@tikzpicture}%
  \BODY
}
\makeatother
\usetikzlibrary{shapes.geometric,arrows}
\tikzstyle{decision} = [diamond, draw, fill=red!20, 
    text width=1cm, text badly centered, minimum height=1cm, minimum width=2cm]
\tikzstyle{block} = [rectangle, draw=black, fill=orange!20, text centered, minimum height=0.5cm, minimum width=2.5cm]
\tikzstyle{signal} = [rectangle, rounded corners, draw=black, fill=blue!20, 
    text width=1cm, text centered, minimum height=1cm, minimum width=2cm]
\tikzstyle{arrow} = [thick,->,>=stealth]

\usepackage{empheq}

\newcommand*\widefbox[1]{\fbox{\hspace{0em}#1\hspace{0em}}}

\usepackage{hyperref}
\hypersetup{colorlinks=false,pdfborder={0 0 0}}

\begin{document}

\title{Adaptive Dereverberation, Noise and Interferer Reduction Using Sparse Weighted Linearly Constrained Minimum Power Beamforming\\
\thanks{This work was funded by the Deutsche Forschungsgemeinschaft (DFG, German Research Foundation) -- Project ID 390895286 -- EXC 2177/1.}
}

\author{\IEEEauthorblockN{Henri Gode and Simon Doclo}
\IEEEauthorblockA{\textit{Department of Medical Physics and Acoustics and Cluster of Excellence Hearing4all},
\textit{University of Oldenburg}, Germany \\
\{henri.gode, simon.doclo\}@uni-oldenburg.de}
}

\maketitle

\begin{abstract}
Interfering sources, background noise and reverberation degrade speech quality and intelligibility in hearing aid applications. In this paper, we present an adaptive algorithm aiming at dereverberation, noise and interferer reduction and preservation of binaural cues based on the \gls{wBLCMP} beamformer. The \gls{wBLCMP} beamformer unifies the multi-channel weighted prediction error method performing dereverberation and the linearly constrained minimum power beamformer performing noise and interferer reduction into a single convolutional beamformer. We propose to adaptively compute the optimal filter by incorporating an exponential window into a sparsity-promoting lp-norm cost function, which enables to track a moving target speaker. Simulation results with successive target speakers at different positions show that the proposed adaptive version of the \gls{wBLCMP} beamformer outperforms a non-adaptive version in terms of objective speech enhancement performance measures. 

\end{abstract}


\glsresetall

\section{Introduction}

In many hands-free speech communication systems such as hearing aids, mobile phones and smart speakers, interfering sounds, ambient noise and reverberation may degrade the speech quality and intelligibility of the recorded microphone signals~\cite{beutelmann_prediction_2006}. 
To enhance speech quality and intelligibility, many multi-microphone speech enhancement methods aiming at noise and interferer reduction and dereverberation have been proposed in the last decades~\cite{doclo_multichannel_2015, vincent_audio_2018}. For many of these methods, both non-adaptive versions with time-invariant parameters as well as adaptive versions with time-varying parameters exist. When considering binaural hearing aids, it is often desired to preserve the binaural cues, which provide spatial awareness of the acoustic scene for the listener~\cite{lavandier_speech_2008}.

A commonly used multi-microphone noise reduction method is the \gls{MPDR} beamformer~\cite{veen_beamforming_1988, cox_robust_1987, gannot_consolidated_2017}, which aims at minimizing the output power while leaving the desired speech component undistorted. The \gls{LCMP} beamformer generalizes the \gls{MPDR} beamformer, providing the possibility of multiple linear constraints, e.g., to perform controlled reduction of the interfering sources~\cite{veen_beamforming_1988, hadad_binaural_2016,goessling_blcmv}. Often the constraints are formulated in terms of the \glspl{RTF} vectors of the target speaker and interfering sources~\cite{reuven_dual-source_2008, markovich_multichannel_2009}. 


To achieve dereverberation, the \gls{WPE} method~\cite{nakatani_speech_2010} and its generalization using sparse priors ~\cite{jukic_multi-channel_2015, jukic_group_2015} are commonly employed. 
\gls{WPE} uses a convolutional filter, applied to a number of past frames in the \gls{STFT} domain, to estimate and subtract the late reverberation component.
Since the \gls{WPE} cost function does not have an analytic solution, it has been proposed to use iterative alternating optimization schemes. In~\cite{yoshioka_adaptive_2009 ,jukic_adaptive_2017} adaptive versions of the \gls{WPE} algorithm have been proposed, e.g., by incorporating an exponential window into the cost function and incorporating an additional constraint to prevent overestimation of the late reverberation~\cite{jukic_adaptive_2017}.

Aiming at joint dereverberation and noise reduction, it has been proposed to perform \gls{MIMO}-\gls{WPE} as a preprocessing stage before \gls{MPDR} beamforming, in a cascade system~\cite{delcroix_strategies_2015}. By unifying the optimization of the convolutional \gls{WPE} filter and the \gls{MPDR} beamformer, the so-called \gls{WPD} beamformer~\cite{nakatani_unified_2019} 
and its generalization using sparse priors~\cite{gode2021joint} were shown to outperform cascade systems. The unified \gls{WPD} beamformer 
is optimized similarly to the \gls{WPE} filter with an additional distortionless constraint using the \glspl{RTF} of the target speaker.
In~\cite{nakatani_simultaneous_2019} two adaptive versions of the \gls{WPD} algorithm have been proposed.

Aiming at joint dereverberation, reduction of interfering sources and noise and preservation of the binaural cues of all sources, the \gls{wBLCMP} beamformer in~\cite{aroudi_cognitive-driven_2020} generalizes the \gls{WPD} beamformer by unifying the optimization of the convolutional \gls{WPE} filter and the \gls{LCMP} beamformer. 
Similarly to~\cite{jukic_adaptive_2017, nakatani_simultaneous_2019}, in this paper, we derive an adaptive version 
by incorporating an exponential window into the cost function, which enables tracking of a moving target speaker. In addition, similarly to~\cite{gode2021joint}, we explicitly control the sparsity of the \gls{STFT} coefficients by using an $\ell_p$-norm cost function. For a complex acoustic scenario featuring a target speaker which suddenly switches position, an interfering source at a fixed position and diffuse babble noise, simulation results show that the adaptive version of the \gls{wBLCMP} beamformer clearly outperforms its non-adaptive version in terms of objective speech enhancement performance measures and \gls{RTF} vector estimation accuracy.


\section{Signal Model}
\label{sec:SignalModel}
We consider $J$ acoustic sources captured by a binaural microphone array setup with $\nicefrac{M}{2}$ microphones on each of two head-worn hearing devices (e.g. left and right hearing aid)  in a noisy and reverberant acoustic environment (with $J < M$). Without loss of generality, the first source ($j=1$) is considered to be the target speaker and the remaining $J-1$ sources are considered to be interfering sources. The \gls{STFT} coefficients of the microphone signals at time frame $t$ are denoted as
\begin{align}
    \mathbf{y}_t = \begin{bmatrix} y_{1,t} & \cdots & y_{M,t} \end{bmatrix}^\mathrm{T} \in \mathbb{C}^{M\times 1},
    \label{eq:signal_vec}
\end{align}
with $\left(\cdot\right)^{\mathrm{T}}$ denoting the transpose operator. In \eqref{eq:signal_vec} the frequency index has been omitted since it is assumed that each frequency subband is independent and hence can be processed individually. 
Similarly to~\cite{jukic_multi-channel_2015, nakatani_unified_2019, nakatani_simultaneous_2019, aroudi_cognitive-driven_2020, gode2021joint}, the multi-channel microphone signal $\mathbf{y}_t$ in \eqref{eq:signal_vec} is modeled as the sum of each source signal $s_{j,t}$ convolved with its possibly time-varying multi-channel \gls{CTF} matrix $\mathbf{A}_{j,t}= \begin{bmatrix} \mathbf{a}_{j,t,0} & \cdots & \mathbf{a}_{j,t,L_a-1} \end{bmatrix} \in \mathbb{C}^{M\times L_a}$ plus background noise $\mathbf{n}_t \in \mathbb{C}^{M\times 1}$, i.e.
\begin{align}
    \mathbf{y}_t = \sum_{j=1}^{J}\sum_{l=0}^{L_a-1} \mathbf{a}_{j,t,l} s_{j,t-l} \ + \ \mathbf{n}_t,
    \label{eq:signal_model}
\end{align}
where $L_a$ denotes the number of taps of the \glspl{CTF}. By splitting the \glspl{CTF} into the early reflections and late reverberation using the integer parameter $\tau$, the reverberant signal for the $j$-th source can be decomposed into its direct component $\mathbf{d}_{j,t} \in \mathbb{C}^{M\times 1}$ (including early reflections)
and its late reverberation component $\mathbf{r}_{j,t} \in \mathbb{C}^{M\times 1}$, i.e.
\begin{align}
     \mathbf{y}_t =  \sum_{j=1}^{J}\underbrace{\sum_{l=0}^{\tau-1} \mathbf{a}_{j,t,l} s_{j,t-l}}_{\coloneqq\mathbf{d}_{j,t}}  \ + \  \sum_{j=1}^{J}\underbrace{\sum_{l=\tau}^{L_a-1} \mathbf{a}_{j,t,l} s_{j,t-l}}_{\coloneqq\mathbf{r}_{j,t}}  \ + \  \mathbf{n}_t.
    \label{eq:signal_model_seperated}
\end{align}
The direct component for the $j$-th source $\mathbf{d}_{j,t}$ can be approximated using the \gls{MTF} vector $\mathbf{v}_{j,t} \in \mathbb{C}^{M\times 1}$ as~\cite{avargel_multiplicative_2007}
\begin{multline}
    \mathbf{d}_{j,t} \approx \mathbf{v}_{j,t} s_{j,t} = \mathbf{\Tilde{v}}_{j,m,t} d_{j,m,t}, \quad m \in \{1,...,M\}, 
    \label{eq:dsigmodel}
\end{multline}
where $d_{j,m,t}$
denotes the direct component of the $j$-th source in the reference microphone $m$ at time frame $t$.
The vector
\begin{align}
    \mathbf{\Tilde{v}}_{j,m,t} = \mathbf{v}_{j,t} / v_{j,m,t} \in \mathbb{C}^{M\times 1}
    \label{eq:RTF_defined}
\end{align}
denotes the possibly time-varying \gls{RTF} vector for the $j$-th source, where $v_{j,m,t}$ is the $m$-th entry of $\mathbf{v}_{j,t}$. 

\section{Sparse wBLCMP Filter}
To obtain an estimate 
of the direct target speech component $d_{1,\nu,t}$ in the left and right 
reference microphone denoted by $m = \nu \in \{L,R\}$, it has been proposed in~\cite{nakatani_unified_2019, nakatani_simultaneous_2019, gode2021joint, aroudi_cognitive-driven_2020} to apply a convolutional filter $\mathbf{\Bar{h}}_{\nu,t}\in\mathbb{C}^{M\left(L_{h}-\tau+1\right)\times 1}$ to the stacked noisy \gls{STFT} vector $\mathbf{\Bar{y}}_{t}$, i.e.
\begin{align}
    \hat{d}_{1,\nu,t} = {\mathbf{\Bar{h}}_{\nu,t}}^{\mathrm{H}}\mathbf{\Bar{y}}_{t},
    \label{eq:WPD_Filter}
\end{align}
where $\left(\cdot\right)^{\mathrm{H}}$ denotes the conjugate transpose operator and the stacked noisy \gls{STFT} vector $\mathbf{\Bar{y}}_{t}$ is defined as
\begin{align}
    \label{eq:stacked_signal}
    \mathbf{\Bar{y}}_{t} &= \begin{bmatrix} \mathbf{y}^{\mathrm{T}}_{t} & | & \mathbf{y}^{\mathrm{T}}_{t-\tau} & \cdots & \mathbf{y}^{\mathrm{T}}_{t-L_{h}+1} \end{bmatrix}^{\mathrm{T}} \in\mathbb{C}^{M\left(L_{h}-\tau+1\right)\times 1},
\end{align}
where $L_{h}$ denotes the filter length.
It should be noted that the vector $\mathbf{\Bar{y}}_{t}$ only includes a subset of the $L_h$ most recent frames, i.e. it includes the current frame but excludes the preceding $\tau-1$ frames, aiming at preserving the early reflections.
\subsection{Non-Adaptive Version}
\label{sec:non_adaptive}
By assuming that all \glspl{CTF} and \glspl{MTF} and the convolutional filter $\mathbf{\Bar{h}}_{\nu,t}$ do not change over time, i.e. $\mathbf{\Bar{h}}_{\nu,t} = \mathbf{\Bar{h}}_{\nu}$ for all time frames $t\in\{1,\ldots,T\}$, a non-adaptive version of the \gls{wBLCMP} beamformer aiming at joint dereverberation, noise and interferer reduction 
has been derived in~\cite{aroudi_cognitive-driven_2020}.
In~\cite{aroudi_cognitive-driven_2020}, assuming that the direct component of the target speaker follows a zero mean complex Gaussian distribution with a time-varying variance $\lambda_n=\abs{d_{1,\nu,n}}^{2}$, the convolutional filter in \eqref{eq:WPD_Filter} is computed by minimizing the negative log-likelihood function
\begin{align}
    \argmin_{\mathbf{\Bar{h}}_{\nu}} \sum_{n=1}^{T} \ln{\lambda_n}+ \frac{\abs{\hat{d}_{1,\nu,n}}^{2}}{\lambda_n} = \sum_{n=1}^{T}\ln{\lambda_n}+\frac{\abs{{\mathbf{\Bar{h}}_{\nu}}^{\mathrm{H}}\mathbf{\Bar{y}}_{n}}^{2}}{\lambda_n},
    \label{eq:WPD_l0_norm_CostFun}
\end{align}
subject to a linear constraint for each source using their \glspl{RTF} defined in \eqref{eq:RTF_defined}, i.e.
\begin{align}
    \mathbf{\Bar{h}}_{\nu}^{\mathrm{H}}\mathbf{\Bar{v}}_{j,\nu} =& \beta_{j}\quad\quad\quad\quad\quad\quad\forall j \in \{1,\ldots,J\}
    \label{eq:linear_constraints} \\
    \mathbf{\Bar{v}}_{j,\nu} =& \begin{bmatrix} \mathbf{\Tilde{v}}_{j,\nu}^{\mathrm{T}} & \mathbf{0}^{\mathrm{T}} \end{bmatrix}^{\mathrm{T}},
\end{align}
where $\mathbf{0}$ denotes a vector containing $M\left(L_h-\tau\right)$ zeros and $\beta_{j}$ denotes a scaling factor for the direct component of the $j$-th source. The scaling factor $\beta_1$ is usually set to 1, corresponding to a distortionless constraint for the target speaker, whereas all other scaling factors are usually chosen to be close to 0, aiming at suppressing the interfering sources.

In this paper, we aim at explicitly taking into account that the \gls{STFT} coefficients of the direct target speech component are sparser than the \gls{STFT} coefficients of the noisy reverberant mixture recorded by the microphones~\cite{jukic_multi-channel_2015}. 
Hence, similarly to the \gls{WPE} variant in~\cite{jukic_group_2015} and the \gls{WPD} variant in~\cite{gode2021joint}, we propose to minimize the convolutional filter in \eqref{eq:WPD_Filter} using an $\ell_p$-norm cost function instead of \eqref{eq:WPD_l0_norm_CostFun}, i.e.
\begin{align}
    \argmin_{\mathbf{\Bar{h}}_{\nu}} \sum_{n=1}^{T}\abs{\hat{d}_{1,\nu,n}}^p = \sum_{n=1}^{T}\abs{{\mathbf{\Bar{h}}_{\nu}}^{\mathrm{H}}\mathbf{\Bar{y}}_{n}}^p
    \label{eq:WPD_lp_norm_CostFun}
\end{align}
where $p \in (0,2] $ denotes the so-called shape parameter. This parameter determines the sparsity of the cost function, where small values of $p$ promote sparsity. It should be noted that for $0 < p < 1$ this cost function is non-convex. 

\subsection{Adaptive Version}
\label{sec:adaptive}
To deal with time-varying acoustic scenarios, e.g. moving sources, in this paper we derive an adaptive version of the \gls{wBLCMP} beamformer. Similarly as in~\cite{jukic_adaptive_2017, nakatani_simultaneous_2019}, we propose to incorporate an exponential window into the cost function in \eqref{eq:WPD_lp_norm_CostFun}. The resulting minimization problem for each time frame $t$ is given by \begin{subequations}
\begin{empheq}[box=\widefbox]{align}
    \label{eq:adaptive_CostFun}
   &\argmin_{\mathbf{\Bar{h}}_{\nu,t}} \sum_{n=1}^{t} \gamma^{t-n}\abs{\hat{d}_{1,\nu,n}}^p =  \sum_{n=1}^{t}\gamma^{t-n}\abs{{\mathbf{\Bar{h}}_{\nu,t}}^{\mathrm{H}}\mathbf{\Bar{y}}_{n}}^p\\ &\mathrm{s.t.}\quad \mathbf{\Bar{h}}_{\nu,t}^{\mathrm{H}}\mathbf{\Bar{v}}_{j,\nu,t} = \beta_{j}\quad\forall j \in \{1,\ldots,J\},
   \label{eq:adaptive_constraints}
\end{empheq}
\end{subequations}
where the smoothing parameter $\gamma\in (0,1]$ allows adaptation to possibly time-varying \glspl{CTF} and \glspl{MTF}. Note that the cost function in \eqref{eq:adaptive_CostFun} reduces to the cost function in \eqref{eq:WPD_lp_norm_CostFun} for $\gamma = 1$ and $t=T$. Therefore, the following derivations based on the adaptive cost function in \eqref{eq:adaptive_CostFun} for the adaptive version also hold for the cost function in \eqref{eq:WPD_lp_norm_CostFun} for the non-adaptive version.

\subsection{Filter Optimization}
\label{sec:alternating_opt}
Similarly as in~\cite{jukic_adaptive_2017, gode2021joint}, we propose to use an \gls{IRLS} procedure to minimize the cost function in~\eqref{eq:adaptive_CostFun} subject to the constraints in~\eqref{eq:adaptive_constraints}. The basic idea is to replace the non-convex $\ell_p$-norm minimization problem with a series of convex $\ell_2$-norm minimization subproblems, which have an analytic solution. In this paper we used only the first iteration of \gls{IRLS}, since preliminary results indicated sufficient convergence. 
\vspace{3mm} 

\subsubsection{Constrained \ensuremath{\ell_2}-Norm Subproblem Minimization}
\label{sec:_l2_norm_subproblem}
\phantom{test} \\ In each frame, the non-convex cost function in \eqref{eq:adaptive_CostFun} is replaced with a convex weighted $\ell_{2}$-norm cost function, i.e.
\begin{align}
    \argmin_{\mathbf{\Bar{h}}_{\nu,t}} \sum_{n=1}^{t} \gamma^{t-n}w_{n}\abs{\hat{d}_{1,\nu,n}}^2 =  \sum_{n=1}^{T}\gamma^{t-n}w_{n}\abs{{\mathbf{\Bar{h}}_{\nu,t}}^{\mathrm{H}}\mathbf{\Bar{y}}_{n}}^2
    \label{eq:WPD_l2_norm_CostFun1}
\end{align}
where the weights $w_{n}$ are real-valued and positive. The filter minimizing \eqref{eq:WPD_l2_norm_CostFun1} subject to the linear constraints in \eqref{eq:adaptive_constraints} is equal to
\begin{align}
    \boxed{  \mathbf{\Bar{h}}_{\nu,t} = \Bar{\mathbf{R}}_{y,t}^{-1}\Bar{\mathbf{C}}_{t}\left(\Bar{\mathbf{C}}_{t}^{\mathrm{H}}\Bar{\mathbf{R}}_{y,t}^{-1}\Bar{\mathbf{C}}_{t}\right)^{-1} \mathbf{B}\Bar{\mathbf{C}}_{t}^{\mathrm{H}}\mathbf{e}_{\nu}},
    \label{eq:WPD_solution}
\end{align} 
where
\begin{align}
    \mathbf{\Bar{R}}_{y,t} = \sum_{n=1}^{t} \gamma^{t-n} w_{n} \mathbf{\Bar{y}}_n\mathbf{\Bar{y}}_n^{\mathrm{H}}
    \label{eq:conv_cov_mat}
\end{align}
denotes the weighted noisy \gls{STCM} of the stacked microphone signals, $\Bar{\mathbf{C}}_{t} =\begin{bmatrix} \mathbf{\Bar{v}}_{1,\nu,t} & \cdots & \mathbf{\Bar{v}}_{J,\nu,t} \end{bmatrix} $ denotes the constraint matrix containing the \gls{RTF} vectors for all sources, $\mathbf{B} = \mathrm{diag}\left(\begin{bmatrix} \beta_{1} & \cdots & \beta_{J} \end{bmatrix}^{\mathrm{T}}\right)$ denotes the diagonal scaling matrix containing the scaling factors for all sources, and $\mathbf{e}_{\nu}$ is a selection vector with its entry corresponding to the left or right reference microphone equal to $1$ and all other entries equal to $0$. Assuming that the weights $w_{n}$ of past frames $n\in\{1,\ldots,t-1\}$ are well estimated during processing of these past frames, the weighted noisy \gls{STCM} $\mathbf{\Bar{R}}_{y,t}$ in \eqref{eq:conv_cov_mat} can be effectively computed by an recursive update in each frame, i.e. $\mathbf{\Bar{R}}_{y,t} = \gamma \mathbf{\Bar{R}}_{y,t-1} + 
    w_{t} \mathbf{\Bar{y}}_t\mathbf{\Bar{y}}_t^{\mathrm{H}}$.
However, since only the inverse of the weighted noisy \gls{STCM} is required in \eqref{eq:WPD_solution} it is more effective to use an update formula for $\mathbf{\Bar{R}}_{y,t}^{-1}$ based on the Woodbury matrix identity, i.e.
\begin{align}
    \mathbf{\Bar{R}}_{y,t}^{-1} = \frac{1}{\gamma}\left(\mathbf{\Bar{R}}_{y,t-1}^{-1} - \frac{w_{t}\mathbf{\Bar{R}}_{y,t-1}^{-1} \mathbf{\bar{y}}_{t}\mathbf{\bar{y}}_{t}^{\mathrm{H}} \mathbf{\Bar{R}}_{y,t-1}^{-1}}{\gamma + w_{t}\mathbf{\bar{y}}_{t}^{\mathrm{H}}\mathbf{\Bar{R}}_{y,t-1}^{-1} \mathbf{\bar{y}}_{t}} \right)
    \label{eq:woodbury}
\end{align}


\subsubsection{Weight Estimation}
\label{sec:_EST_Weights}
\phantom{test} \\ Similarly as in~\cite{jukic_multi-channel_2015, gode2021joint}, in each frame $t$ the weight $w_{t}$ in \eqref{eq:woodbury} is estimated as
\begin{align}
    w_{t} = \left( \sum_{\nu} \abs{\hat{d}_{1,\nu,t}}^{2}  \right)^{\nicefrac{p}{2}-1} = \left( \sum_{\nu} \abs{{\mathbf{\Bar{h}}_{\nu,t}}^{\mathrm{H}}\mathbf{\Bar{y}}_{t}}^{2}  \right)^{\nicefrac{p}{2}-1},
    \label{eq:update_weights}
\end{align}
such that \eqref{eq:WPD_l2_norm_CostFun1} is a first-order approximation of \eqref{eq:adaptive_CostFun}. Note that the shape parameter $p$ only affects the weight update in \eqref{eq:update_weights} of the algorithm, where it is possible to set $p=0$.

\begin{figure}[t!]
    \centering
    \begin{tikzpicture}[node distance=1.1cm]
            \node (wBLCMP) [block] {4. \gls{wBLCMP} Beamformer};
            \node (Estimation) [block, below of=wBLCMP, xshift=0cm] {3. Estimate $\mathbf{\Bar{R}}_{y,t}$ and $\mathbf{\Bar{h}}_{\nu\,t}$};
            \node (RTF) [block, below of=Estimation, xshift=0cm] {2. Estimate \glspl{RTF} $\mathbf{\Bar{C}}_{t}$};
            \node (WPE) [block, left of=RTF, xshift=-2.35cm] {1. \gls{MIMO}-\gls{WPE}};
            \node (sig) [left of=WPE, yshift=2.2cm, xshift=-0.9cm] {$\mathbf{y}_t$};
            \node (outsig) [right of=wBLCMP, xshift=1.7cm] {$\hat{d}_{1,\nu,t}$};
            
            \draw [arrow] (wBLCMP) -- (outsig);
            \draw [arrow] (sig) -- (wBLCMP);
            \draw [arrow] (sig) -- +(0.45,0) |- (Estimation);
            \draw [arrow] (sig) -- +(0.45,0) |- (WPE);
            \draw [arrow] (WPE) -- node[anchor=south] {$\mathbf{z}_t$} (RTF);
            \draw [arrow] (RTF) -- node[anchor=west] {$\mathbf{\Bar{C}}_{t}$} (Estimation);
            \draw [arrow] (Estimation) -- node[anchor=west] {$\mathbf{\Bar{h}}_{\nu\,t}$} (wBLCMP);

    \end{tikzpicture}
    \vspace*{-5mm}
    \caption{Block diagram of the proposed adaptive \gls{wBLCMP} algorithm, incorporating a \gls{MIMO}-\gls{WPE} preprocessing stage for estimating the \glspl{RTF}.}
    \label{fig:Flow_Chart_MIMO_WPD}
\end{figure}
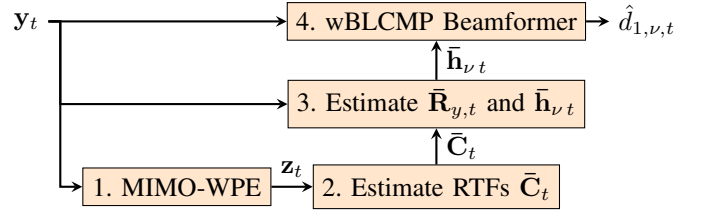

\subsection{RTF Estimation}

The \gls{wBLCMP} beamformer in~\eqref{eq:WPD_solution} requires estimates of the \glspl{RTF} for each source, which can be obtained using the covariance whitening method~\cite{markovich_multichannel_2009, serizel_low-rank_2014}. It has been shown in~\cite{nakatani_simultaneous_2019} that performing \gls{RTF} estimation on multi-channel dereverberated signals $\mathbf{z}_{t}$, obtained by a \gls{MIMO}-\gls{WPE} preprocessing stage, is beneficial, since the \gls{MTF}-based model in \eqref{eq:dsigmodel} assumes short transfer functions for the direct component. The block diagram in Fig.~\ref{fig:Flow_Chart_MIMO_WPD} shows an overview of the complete algorithm. Note that the computation time is not significantly increased by the \gls{MIMO}-\gls{WPE} preprocessing stage, since the \gls{wBLCMP} filter can be effectively computed using the \gls{MIMO}-\gls{WPE} filter, because both are based on the convolutional signal model in \eqref{eq:signal_model} and can be derived using the $\ell_p$-norm cost function in \eqref{eq:adaptive_CostFun} \cite{nakatani_maximum_2019}. The \gls{RTF} vector of the $j$-th source can then be estimated based on the generalized eigenvalue decomposition of the dereverberated covariance matrix $\mathbf{R}_{j,t}$ of that source and the dereverberated covariance matrix $\mathbf{R}_{v,j,t}$ of all other sources and the background noise. Since accurately estimating all of these covariance matrices is far from trivial, in this paper, we will assume oracle knowledge about a noise-only period and a noise-plus-interferer period in the beginning of the signal, which are used to compute fixed covariance matrices of an interfering source and noise. In contrast, the covariance matrix and \gls{RTF} vector of the target are tracked.

\section{Experimental Results}

In this section, we compare the performance of the proposed adaptive version of the \gls{wBLCMP} beamformer (Sec.~\ref{sec:adaptive}) with the non-adaptive version (Sec.~\ref{sec:non_adaptive}) using different shape parameters $p$ for a spatially non-stationary acoustic scenario where the target speaker suddenly switches position.

\subsection{Acoustic Scenario}

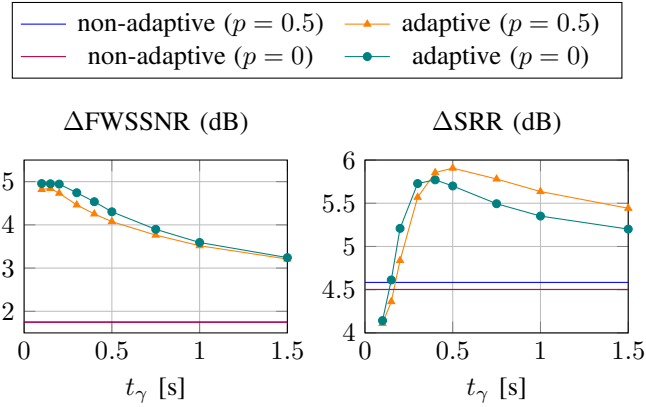
\begin{figure}[t]
    \begin{center}
    \begin{tikzpicture}
\begin{customlegend}[legend columns=2,legend style={align=center,column sep=0.5ex, at={(1,0)}},
        legend entries={non-adaptive ($p=0.5$) ,
                        adaptive ($p=0.5$) ,
                        non-adaptive ($p=0$) ,
                        adaptive ($p=0$) ,
                        }]
        \addlegendimage{mark=, mark size=1.5pt, blue,solid,line legend}
        \addlegendimage{mark=triangle*, mark size=1.5pt, orange}  
        \addlegendimage{mark=, mark size=1.5pt, purple}
        \addlegendimage{mark=*, mark size=1.5pt, teal}
        \end{customlegend}
        \end{tikzpicture}
      \end{center}
    \begin{tikzpicture}
    \matrix{
        \begin{axis}[name=plot7,
                     title={$\Delta$FWSSNR (dB)},
                     xlabel={$t_{\gamma}$ [s]},
                     xmin = 0,
                     xmax = 1.5,
                     ymin = 1.5,
                     ymax = 5.5,
                     grid=major,
                     ytick distance=1,
                     ]

      \addplot [mark=, mark size=1.5pt, blue] 
      coordinates { 
      (0,1.7523) (0.1000,1.7523) (0.1500,1.7523) (0.2000,1.7523) (0.3000,1.7523) (0.4000,1.7523) (0.5000,1.7523) (0.7500,1.7523) (1.0000,1.7523) (1.5000,1.7523) 
      }; 
      
      \addplot [mark=triangle*, mark size=1.5pt, orange] 
      coordinates { 
      (0.1000,4.8182) (0.1500,4.8419) (0.2000,4.7326) (0.3000,4.4609) (0.4000,4.2525) (0.5000,4.0753) (0.7500,3.7628) (1.0000,3.5173) (1.5000,3.2087) 
      }; 
                     
            \addplot [mark=, mark size=1.5pt, purple] 
      coordinates { 
      (0,1.7460) (0.1000,1.7460) (0.1500,1.7460) (0.2000,1.7460) (0.3000,1.7460) (0.4000,1.7460) (0.5000,1.7460) (0.7500,1.7460) (1.0000,1.7460) (1.5000,1.7460) 
      }; 
      
      \addplot [mark=*, mark size=1.5pt, teal] 
      coordinates { 
      (0.1000,4.9556) (0.1500,4.9508) (0.2000,4.9436) (0.3000,4.7432) (0.4000,4.5360) (0.5000,4.3022) (0.7500,3.8968) (1.0000,3.5922) (1.5000,3.2425) 
      }; 
      
      
        \end{axis}
        
       &\begin{axis}[name=plot8,
                     title={$\Delta$SRR (dB)},
                     xlabel={$t_{\gamma}$ [s]},
                     xmin = 0,
                     xmax = 1.5,
                     ymin = 4,
                     ymax = 6,
                     grid=major,
                     ]
                     
      \addplot [mark=, mark size=1.5pt, blue] 
      coordinates { 
      (0,4.5844) (0.1000,4.5844) (0.1500,4.5844) (0.2000,4.5844) (0.3000,4.5844) (0.4000,4.5844) (0.5000,4.5844) (0.7500,4.5844) (1.0000,4.5844) (1.5000,4.5844) 
      }; 
      
      \addplot [mark=triangle*, mark size=1.5pt, orange] 
      coordinates { 
      (0.1000,4.1094) (0.1500,4.3620) (0.2000,4.8378) (0.3000,5.5694) (0.4000,5.8544) (0.5000,5.9065) (0.7500,5.7806) (1.0000,5.6355) (1.5000,5.4422) 
      };
                     
          \addplot [mark=, mark size=1.5pt, purple] 
      coordinates { 
      (0,4.5033) (0.1000,4.5033) (0.1500,4.5033) (0.2000,4.5033) (0.3000,4.5033) (0.4000,4.5033) (0.5000,4.5033) (0.7500,4.5033) (1.0000,4.5033) (1.5000,4.5033) 
      }; 
      
      \addplot [mark=*, mark size=1.5pt, teal] 
      coordinates { 
      (0.1000,4.1426) (0.1500,4.6126) (0.2000,5.2093) (0.3000,5.7281) (0.4000,5.7708) (0.5000,5.7000) (0.7500,5.4946) (1.0000,5.3518) (1.5000,5.2010) 
      }; 
      
      
     
        \end{axis}
        
      
      
      
     
    \\
    };
    \end{tikzpicture}
    \vspace*{-7.5mm}
    \caption{Average FWSSNR and SRR improvement vs. time constant $t_{\gamma}$ for different values of the shape parameter $p$. Note that the non-adaptive method obviously does not have a time constant.}
    \label{fig:Delta_Perf_Lines}
\end{figure}

We considered 2 \gls{BTE} hearing aids with 2 microphones each, mounted on a dummy head located approximately in the center of an acoustic laboratory $(\SI{7}{\m}\times \SI{6}{\m} \times \SI{2.7}{\m})$ with a reverberation time $T_{60} \approx \SI{510}{\milli\s}$.
The acoustic scenario consists of one target speaker (which suddenly switches position), one interfering speaker (at a fixed position) and background noise. The target and interfering speech components at the microphones were generated by convolving clean speech signals with room impulse responses measured from loudspeakers at about $\SI{2}{\m}$ from the dummy head. The target speaker at position 1 ($\SI{0}{\degree}$, front of dummy head) is a male speaker which is active in the interval $\left[\SI{2}{\s},\SI{20.4}{\s}\right]$, whereas the target speaker at position 2 (\SI{90}{\degree}, right of dummy head) is a female speaker which is active in the interval $\left[\SI{20.4}{\s},\SI{39}{\s}\right]$. The interfering speaker is a male speaker which is located at $\SI{-120}{\degree}$ and is active in the interval $\left[\SI{1}{\s},\SI{39}{\s}\right]$.
Quasi-diffuse babble noise, which is constantly active, was generated by playing back cafeteria noise 
using 4 loudspeakers facing the corners of the laboratory.
The noisy mixture is constructed at a broadband \gls{SNR} of $\SI{0}{\dB}$ and a broadband \gls{SIR} of $\SI{0}{\dB}$ for both target positions. Note that there is a noise-only period in the $1^{\mathrm{st}}$ second and a noise-plus-interferer period in the $2^{\mathrm{nd}}$ second.
The sampling frequency was equal to \SI{16}{\kilo\hertz}.

\subsection{Algorithm Settings}
\label{sec:settings}
We applied the \gls{wBLCMP} beamformer within an \gls{STFT} framework with a frame length of $\SI{32}{\milli\s}$, a frame shift of $t_s = \SI{16}{\milli\s}$ and a sqrt-Hann window for analysis and synthesis. We compared the performance of two shape parameters $p=\{0,0.5\}$, since it has been shown in~\cite{jukic_multi-channel_2015} that a shape parameter of $p=0.5$ can be beneficial. The filter length $L_h$ in \eqref{eq:stacked_signal} was set to 16 frames corresponding to $\SI{256}{\milli\s}$ covering about half of the $T_{60}$. The prediction delay $\tau$ was set to 3 frames corresponding to $\SI{48}{\milli\s}$ aiming at preserving the early reflections.
The scaling factors of the target speaker and the interfering source in \eqref{eq:linear_constraints} were set to $\beta_1 = \SI{0}{dB}$ and $\beta_2 = \SI{-20}{\dB}$, respectively. Since preliminary results indicated reasonable convergence after the initial iteration of the alternating optimization described in Sec.~\ref{sec:alternating_opt}, we chose to stop after the first iteration to reduce computational cost. For the adaptive versions, different time constants are evaluated between $t_{\gamma} = \left[\SI{100}{\milli\s}, \SI{1500}{\milli\s}\right]$. The smoothing parameter $\gamma$ can be computed using the time constant as $\gamma = e^{\nicefrac{-t_{s}}{t_{\gamma}}}$. The noise-plus-interferer covariance matrix $\mathbf{R}_{v,2}$ and \gls{RTF} vector of the interfering source $\mathbf{\Tilde{v}}_{2,\nu}$ are fixed after the first $\SI{2}{\s}$, whereas the covariance matrix and \gls{RTF} vector of the target speaker are adaptively tracked.


\begin{figure}[t]
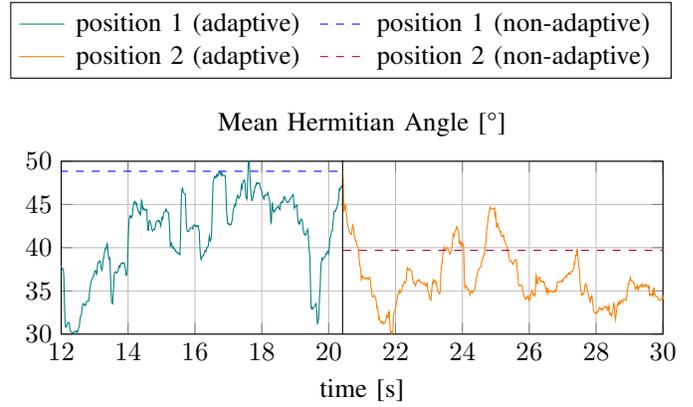

    \begin{center}

    \vspace*{-7.5mm}
    \caption{Mean Hermitian angle between the oracle \gls{RTF} vector of the active target speaker and the estimated target \gls{RTF} vector within the \gls{wBLCMP} algorithm ($p=0.5$) over time for a time constant $t_{\gamma} = \SI{400}{\milli\s}$. The switch of target speaker occurs at approximately $\SI{20.4}{\s}$. Note that the non-adaptive version only provides one constant \gls{RTF} vector estimate for the whole signal.}
    \label{fig:Herm_angle}
\end{figure}

\subsection{Objective Speech Enhancement Measures}

As objective performance measures we used the \gls{FWSSNR}~\cite{hu_evaluation_2008}, and the \gls{SRR}~\cite{naylor_signal-based_2010} averaged across the left and right output signal.
As reference signal for \gls{FWSSNR} and \gls{SRR} we used the direct target speech component including early reflections (first $\SI{50}{\milli\s}$ of the \glspl{RIR}) at the reference microphones.

In addition, we evaluate the \gls{RTF} vector estimation accuracy based on the  Hermitian angle
\begin{align}
\varphi = \mathrm{acos} \left( \frac{\abs{\mathbf{\hat{\tilde{v}}}_{j,t}^{\mathrm{H}}\mathbf{\tilde{v}}_{j,t}}}{\norm{\mathbf{\hat{\tilde{v}}}_{j,t}}\norm{\mathbf{\tilde{v}}_{j,t}}} \right)
\end{align}
between the estimated \gls{RTF} vector $\mathbf{\hat{\tilde{v}}}_{j,t}$ of the target speaker and the oracle \gls{RTF} vector $\mathbf{\tilde{v}}_{j,t}$ averaged across frequency bands. The Hermitian angle $\varphi$ is a scale-invariant error measure for complex vectors, with lower values indicating smaller errors. The oracle \gls{RTF} vectors are computed as the principal eigenvector of the covariance matrices of a white noise signal convolved with the early part ($\SI{50}{\milli\s}$) of the respective multi-channel \glspl{RIR} of the target speaker. Note that for each target speaker position there is a unique oracle \gls{RTF} vector. 

\subsection{Results}

Fig.~\ref{fig:Delta_Perf_Lines} compares the \gls{FWSSNR} and \gls{SRR} improvements (difference between scores for input and output signals) for different time constants $t_{\gamma}$ of the adaptive and non-adaptive version of the \gls{wBLCMP} beamformer using two different shape parameters $p=\{0,0.5\}$. It can be clearly observed that for the considered switching-target scenario the adaptive version of the \gls{wBLCMP} beamformer outperforms the non-adaptive version in both performance measures for almost all time constants. The best \gls{SRR} improvement is obtained using a time constant of roughly $t_{\gamma} = \SI{450}{\milli\s}$, whereas the \gls{FWSSNR} improvement is higher for shorter time constants. Using the shape parameter $p=0.5$ yields better \gls{SRR} improvements especially for larger time constants, whereas using the shape parameter $p=0$, corresponding to the conventional cost function in \eqref{eq:WPD_l0_norm_CostFun}, yields slightly better \gls{FWSSNR} improvements.

For the adaptive and the non-adaptive version Fig.~\ref{fig:Herm_angle} shows the average Hermitian angle between the oracle \gls{RTF} vector of the active target speaker and the estimated target \gls{RTF} vector. Note that the non-adaptive version only provides one \gls{RTF} vector estimate for the whole signal in contrast to the adaptive version which estimates the \gls{RTF} vector of the target speaker in each time frame. It can be observed that the adaptive \gls{wBLCMP} beamformer outperforms the non-adaptive version in almost all time frames in terms of \gls{RTF} vector estimation accuracy.

\section{Conclusion}

In this paper, we derived an adaptive version of the \gls{wBLCMP} beamformer capable of tracking a moving target speaker in a noisy environment with interfering sources. In addition we generalized the conventional method using sparse priors. The evaluation in terms of objective performance measures clearly shows that the adaptive version outperforms the non-adaptive version in the considered acoustic scenario. This can be explained partly by the ability to track the time-varying \gls{RTF} vector and covariance matrix of a moving target speaker. 


\bibliographystyle{IEEEtran}

\bibliography{mybib}


\end{document}